\RequirePackage{fix-cm}
\documentclass[twocolumn]{svjour3}          
\smartqed  
\usepackage{hyperref, amssymb, amsmath}
\usepackage{graphicx}
\usepackage{xcolor}
\usepackage{wasysym}

\journalname{Statistics and Computing}
\begin{document}

\renewcommand{\epsilon}{\varepsilon}
\renewcommand{\phi}{\varphi}
\definecolor{A}{HTML}{F8766D}
\definecolor{B}{HTML}{00BFC4}

\title{Nonparametric likelihood based estimation of
  linear filters for point processes
}

\titlerunning{Point process likelihood estimation of linear filters}        

\author{Niels Richard Hansen}


\institute{Niels Richard Hansen \at
              University of Copenhagen,
              Department of Mathematical Sciences, Universitetsparken 5, 2100 Copenhagen, Denmark. \\
              Tel.: +45 - 35 32 07 83 \\
              \email{Niels.R.Hansen@math.ku.dk}           
}

\date{}

\maketitle

\begin{abstract}

  We consider models for multivariate point processes where the
  intensity is given nonparametrically in terms of functions in a
  reproducing kernel Hilbert space. The likelihood function involves a
  time integral and is consequently not given in terms of a finite
  number of kernel evaluations. The main result is a representation of
  the gradient of the log-likelihood, which we use to derive computable
  approximations of the log-likelihood and the gradient by time
  discretization. These approximations are then used to minimize the approximate
  penalized log-likelihood. For time and memory efficiency the
  implementation relies crucially on the use of sparse matrices. As an
  illustration we consider neuron network modeling,
  and we use this example to investigate how the computational costs of
  the approximations depend on the resolution of the time discretization. 
The implementation is available in the R package \texttt{ppstat}.

\keywords{Multivariate point processes \and Penalization \and Reproducing kernel Hilbert
  spaces \and \texttt{ppstat}}

\end{abstract}

\section{Introduction}
\label{intro}

Reproducing kernel Hilbert spaces have become widely used in 
statistics and machine learning,
\cite{Bishop:2006}, \cite{Hastie:2009}, \cite{Scholkopf:2001}, where they provide a 
means for nonparametric estimation of nonlinear
functional relations. They have a long history in the statistical
literature, with noteworthy relations to splines and other basis
expansion techniques, as well as Gaussian process theory, see
\cite{Berlinet:2004}, \cite{Wahba:1990}. A typical application is
to estimation of a mean value that depends on one or more predictor
variables. As a function of the predictor variables the mean is
assumed to be in a Hilbert space, and the estimator is obtained by penalized
estimation in the linear Hilbert space -- using the Hilbert space norm 
for penalization. If we use the squared error loss, the so-called
representer theorem states
that for a reproducing kernel Hilbert space the estimation problem 
is, in fact, a finite dimensional optimization problem. The
optimization problem is given in terms of a finite number of kernel evaluations, see
\cite{Hofmann:2008} for a recent review. The representer theorem holds
for any loss function, which is a function of a finite number of continuous
linear functionals. 
 
In this paper we show how to use
reproducing kernel Hilbert space techniques for  
nonparametric point process modeling of e.g. neuron network activity.
 A network of neurons is a prime example of an interacting
dynamical system, and the characterization and modeling of the network
activity is a central scientific challenge, see
e.g. \cite{Pillow:2008}. Data consist of a 
collection of spike times, which can be measured simultaneously for
multiple neurons. The spike times are discrete event times and the
appropriate modeling framework is that of multivariate point
processes. From a prediction viewpoint the objective is to predict 
the next spike time of a given
neuron as a function of the history of the 
spike times for all neurons.   

A natural modeling approach is via the conditional intensity, which
specifies how the history affects the immediate intensity -- or rate
-- of the occurrence of another spike. The negative log-likelihood for a point process model is given
directly in terms of the intensity, but the representer theorem, Theorem 9 in
\cite{Hofmann:2008}, does not hold in general, see
\cite{Hansen:2013a}. The reason is that the log-likelihood involves a
time-integral, see (\ref{eq:negloglikelihood}) below, and the log-likelihood is
consequently not a function of a \emph{finite} number of continuous
linear functionals in general. This is the main problem that we address in this
paper. 

To motivate our general nonparametric model class we briefly review the
 classical linear Hawkes process introduced by Hawkes in 1971,
 \cite{Hawkes:1971}. We consider in the following $p$ different
 counting processes of discrete events, e.g. spike times. We 
let $(N^j_t)$ denote the $j$'th of the counting processes, for $j =
 1, \ldots, p$, and we assume first that the intensity of a new event for the $i$'th process
 is $Y^i_t = \sum_{j=1}^p Y^{ij}_t$ where 
\begin{equation} \label{eq:expfilt}
Y_t^{ij} =  \int_0^{t-} e^{\alpha_{ij} (t-s) + \beta_{ij}} \,
\mathrm{d} N_s^j.
\end{equation}
The intensity process $Y^i$ jumps by $e^{\beta_{ij}}$ whenever an
event occurs in the $j$'th process, and the $\alpha_{ij}$-parameters
control the smooth exponential behavior in between jumps. The
intensity specifies the conditional probability of observing
an event immediately after time $t$ in the sense that
$$P(N^i_{t + \delta} - N^i_t = 1 \mid \mathcal{F}_t) \simeq \delta Y_t^i,$$
where $ \mathcal{F}_t$ denotes the history of all events preceding
time $t$, see e.g. \cite{Jacobsen:2006} or
\cite{AndersenBorganGillKeiding:1993}. 
Note the upper integration limit, $t-$, which means that the
integral w.r.t. $N_s^j$ only involves events strictly before $t$. This
is an essential requirement for correct likelihood computations, see
(\ref{eq:negloglikelihood}) below. 

It follows from  (\ref{eq:expfilt}) that if 
$\sigma_j < t$ denotes the last of the $j$'th events before $t$, 
$$Y_t^{ij} = e^{\alpha_{ij} (t - \sigma_j)} Y_{\sigma_j+}^{ij}.$$
This provides an efficient way of computing the intensity process.
In fact, it follows that $(Y_t^{ij})_{i,j=1, \ldots,p}$ 
is a $p^2$-dimensional Markov process, and that there is a 
one-to-one correspondence between this process and the
multivariate counting process $(N_t^1, \ldots, N_t^p)$. 

Our interest is to generalize the model given in terms of
(\ref{eq:expfilt}) to non-exponential integrands, and, in particular,
to allow those integrands to be estimated nonparametrically. A
consequence is that the Markov property will be lost, and that the
intensity computation will be more demanding. 

The integral (\ref{eq:expfilt}) can be understood as a linear filter of
the multivariate counting process $(N_t^1, \ldots, N_t^p)$, and we will
consider the generalization of such linear filters to processes of the form
\begin{equation} \label{eq:filt}
X_t^i =  \sum_{j=1}^p \int_0^{t} g_{ij}(t-s) \,
\mathrm{d} N_s^j
\end{equation}
with $g_{ij}$ general functions in a suitable function space. The
positive $Y^i$ process is a special case with $g_{ij}(s) = e^{\alpha_{ij} s +
  \beta_{ij}}$. If we allow for negative integrands in (\ref{eq:filt}) the linear filter
 can take negative values, in which case it cannot be an intensity. 
We therefore consider transformations of $X_t^i$, such that
the intensity is given by $\phi(X_t^i)$ for a general but fixed
function $\phi : \mathbb{R} \to [0,\infty)$.

In this paper we are particularly concerned with efficient computation
and minimization of the penalized negative
log-likelihood as a function of the non-parametric components
$g_{ij}$, with $g_{ij}$ in a reproducing kernel Hilbert space $\mathcal{H}$. We consider
algorithms for standard quadratic penalization $\sum_{ij} ||g_{ij}||^2$, with
$|| \cdot ||$ the Hilbert space norm on $\mathcal{H}$. We will
throughout assume that the $g_{ij}$-functions are variation
independent, which imply that the computation and minimization of the
joint penalized negative log-likelihood can be split into $p$ separate
minimization problems. To ease notation we will thus subsequently consider the modeling of one
counting process $N_t$ in terms of $N_t^1, \ldots, N_t^p$, where $N_t$ can be
any of the $p$ counting processes. All computations presented were carried out in
R using the \texttt{ppstat} package. It implements the models of
multivariate point processes through linear filters that are treated in
this paper.  


\section{Likelihood computations for point processes specified by linear filters}

We assume that we observe a simple counting process
$(N_s)_{0 \leq s \leq t}$ of discrete events on the time interval $[0,t]$.
The jump times of $N_s$ are denoted $\tau_1 <  \ldots < \tau_{N_t}$.  We
let $\mathcal{H}$ denote a
reproducing kernel Hilbert space of functions on $[0,t]$ with
reproducing kernel $R : [0,t] \times [0,t] \to \mathbb{R}$, and we let $g = (g_1, \ldots, g_p) \in
\mathcal{H}^p$. We assume that $R$ is continuous in which case the
functions in $\mathcal{H}$ are also continuous, see Theorem 17 in
\cite{Berlinet:2004}. With $N^1, \ldots, N^p$ counting processes with
corresponding event times $\sigma_j^i$ we introduce 
\begin{eqnarray*}
X_s(g) & = & \sum_{i=1}^p \int_0^{s-} g_{i}(s - u) \, \mathrm{d}
N^i_u \\ & = & \sum_{i=1}^p \sum_{j: \sigma_j^i < s} g_i(s - \sigma_j^i).
\end{eqnarray*}
As a function of $g$ we note that $X_s : \mathcal{H}^p \to \mathbb{R}$
being a sum of function evaluations is a continuous linear
functional. Indeed, we can express $X_s$ in terms of inner products
with the kernel as  
\begin{equation} \label{eq:rep1}
X_s(g)  =  \sum_{i=1}^p \sum_{j: \sigma_j^i < s} \langle R(s -
\sigma_j^i, \cdot), \ g_i \rangle.
\end{equation}
The process $X_s(g)$ is called the \emph{linear predictor}
process. We consider the model of $(N_s)_{0 \leq s \leq t}$ where the intensity is given as
$\lambda_s(g) = \phi(X_s(g))$ with $\phi : \mathbb{R} \to [0,\infty)$
a known function. The objective is to estimate the
$g_i$-functions in $\mathcal{H}$. In most applications we will 
include a baseline parameter as well, in which case the linear predictor
becomes $\beta_0 + X_s(g)$. In order not to complicate the notation 
unnecessarily we take $\beta_0 = 0$ in the theoretical presentation. 

From Corollary II.7.3 in \cite{AndersenBorganGillKeiding:1993} it
follows that the negative log-likelihood w.r.t. the homogeneous
Poisson process is given as 
\begin{equation} \label{eq:negloglikelihood}
\ell(g) = \int_0^t \phi(X_s(g)) \, \mathrm{d} s - \sum_{k=1}^{N_t}
\log \phi(X_{\tau_k}(g)).
\end{equation}
If $\phi$ is the identity the time integral has a closed form
representation in terms of the antiderivatives of $g_i$, but in
general it has to be computed numerically. We should note that $\ell$
is convex as a function of $g$ if $\phi$ is convex and
log-concave. 

The following proposition gives the gradient of $\ell$ in the
reproducing kernel Hilbert space. This result is central for our development
 and understanding of a practically implementable minimization
algorithm of the penalized negative log-likelihood. 

\begin{proposition} \label{prop:grad} If $\phi$ is continuously differentiable the 
gradient in $\mathcal{H}$ w.r.t. $g_i$ is 
\begin{eqnarray} \nonumber
\nabla_i \ell (g) & = & \sum_{j} \int_{\sigma_j^i}^t \phi'(X_{s}(g))
R(s - \sigma_j^i, \cdot) \, \mathrm{d} s \\
& & -  \label{eq:gradient}
  \sum_{j} \sum_{k: \sigma_j^i < \tau_k} \frac{\phi'(X_{\tau_k}(g))}{\phi(X_{\tau_k}(g))}
  R(\tau_k - \sigma_j^i, \cdot).
\end{eqnarray}
\end{proposition}

The proof of Proposition \ref{prop:grad} is given in Section
\ref{sec:proof}. It is a special case of Proposition 3.6 in
\cite{Hansen:2013a} if $\mathcal{H}$ is a Sobolev space. However, 
since we restrict attention to counting process integrators in this
paper, in contrast to \cite{Hansen:2013a} where more general
integrator processes are allowed, we can give a relatively elementary
proof for $\mathcal{H}$ being any reproducing kernel Hilbert space with a
continuous kernel.

Computations of $\ell$ as well as the gradient involve the
computation of $X_s(g)$. Without further assumptions
a direct computation of $X_s(g)$ on a grid of $n$ time points involves
in the order of $n \sum_{i=1}^p N_t^i$ evaluations of the
$g_i$-functions. In comparison, (\ref{eq:expfilt}) can be computed
recursively with the order of $np$ evaluations of the exponential
function. 

In this paper we consider three techniques for reducing the general
costs of computing $X_s(g)$. 

\begin{itemize}
\item {\bf Bounded memory}. The filter functions $g_i$ are restricted to have
  support in $[0,A]$ for a fixed $A$. 
\item {\bf Preevaluations}. The filter functions are preevalu{-}ated
  on a grid in $[0,A]$.
\item {\bf Basis expansions}. The filter functions are of the form $g
  = \sum_{k} \beta_k B_k$ for fixed basis functions $B_k$, and 
$$X_s(g) = \sum_{k} \beta_k X_s(B_k).$$
The linear filters $X_s(B_k)$ are precomputed. 
\end{itemize}

\section{Time discretization} \label{sec:discrete}

In this section we discuss the time discretizations necessary for the
practical implementation of an optimization algorithm in
$\mathcal{H}$.  We assume that all filter functions $g_i$ have a
prespecified support restricted to $[0, A]$, and that $\mathcal{H}$ is
restricted to be a space of functions with support in $[0, A]$.  We
approximate time integrals by right Riemann sums with functions
evaluated in the grid
$$0 = t_0 < t_1 < \ldots < t_n = t$$
 and corresponding interdistances $\Delta_l = t_l - t_{l-1}$
for $l = 1, \ldots, n$. We will assume that the collection of event
times is a subset of this grid and denote the corresponding subset of
indices by $I_{\text{jump}} \subseteq \{0, \ldots, n\}$.

We need an implementable representation of the linear predictor as
well as the functional gradient given by (\ref{eq:gradient}). A possible representation of $g_i$
itself is via the $N$-dimensional vector $\mathbf{g}_i$ of its evaluations in
a grid 
$$0 = \delta_0 < \delta_1 < \ldots < \delta_N = A,$$ 
that is,
$\mathbf{g}_{ik} = g_i(\delta_k)$ for $k = 0, \ldots, N-1$. We let
$\mathbf{g}$ denote the $N \times p$ matrix with columns $\mathbf{g}_i$'s for $i = 1, \ldots, p$. Define
$$h_{lik}  = \# \{ j \mid \delta_k \leq t_l - \sigma_j^i <
\delta_{k+1}\} 1( \sigma_j^i < t_l)$$ 
as the number of events for $N^i$ in 
\mbox{$(t_l - \delta_{k+1}, t_l - \delta_k]$.} The indicator $1( \sigma_j^i <
t_l)$ ensures that if $ \sigma_j^i = t_l$ then $h_{li0} = 0$, which, in
turn, ensures that the approximation of the linear predictor below does not
anticipate events. It is the intention that the grid is chosen such that the
$h_{lik}$'s take the values 0 and 1 only. The linear predictor
for given $g_i$'s evaluated in the grid points is approximated as
\begin{eqnarray}
\xi_l & := & \sum_{i,k} h_{lik} \mathbf{g}_{ik} \label{eq:linpred}
\\ & \simeq & \sum_{i=1}^p \sum_{j: t_l - A \leq \sigma_j^i <
  t_l} g(t_l - \sigma_j^i) \\ 
& = & \sum_{i=1}^p \int_{t_l - A}^{t_l-} g^i(t_l - u)
\mathrm{d} N^i_u.
\nonumber
\end{eqnarray}
To formally handle the lower limit in
the integral correctly, $h_{li(N-1)}$ should be redefined to be 1 if
$\sigma_j^i = t_l - A$. Such a redefinition will typically have no detectable
consequences, whereas handling the case $\sigma_j^i = t_l$ correctly
is crucial to avoid making the approximation anticipating.  
An approximation of the negative log-likelihood in $g$ is then
obtained as  
\begin{equation} \label{eq:mllapp}
\ell^{\text{approx}}(\mathbf{g}) = \sum_{l} \phi(\xi_l) \Delta_l - \sum_{l \in
  I_{\text{jump}}} \log \phi(\xi_l).
\end{equation}
If we use the same $\delta$-grid for evaluating the kernel $R$, we
get the gradient approximation from Proposition \ref{prop:grad} 
\begin{eqnarray} \nonumber
\nabla_i \ell^{\text{approx}}(\mathbf{g}) & = & \sum_{k} \left(\sum_{l} \phi'(\xi_l) \Delta_l
h_{lik} \right)R(\delta_k,
\cdot) \\ &&  \label{eq:dmllapp1} - \sum_{k} \left(\sum_{l \in I_{\text{jump}}} \frac{\phi'(\xi_l)}{\phi(\xi_l)}
h_{lik} \right) R(\delta_k, \cdot).
\end{eqnarray}
We observe that 
$$\nabla_i \ell^{\text{approx}}(\mathbf{g}) \in \text{span}\{R(\delta_0, \cdot), \ldots, R(\delta_{N-1}, \cdot)\}.$$
The consequence is that any descent algorithm based on $\nabla_i
\ell^{\text{approx}}(\mathbf{g})$ stays in the finite
dimensional subspace spanned by $R(\delta_0, \cdot), \ldots,
R(\delta_{N-1}, \cdot)$ -- if it starts in this subspace. As we show
below, there is a unique element in this subspace with evaluations
$\mathbf{g}_i$, and the discretization effectively restricts  
$g_i$ to be a function in this subspace. 

\subsection{The direct approximation} 

The  $N \times N$ Gram matrix $\mathbb{G}$ is 
given as $\mathbb{G}_{kl} =
  R(\delta_k, \delta_l)$. The vector $\mathbf{g}_i$
 can be identified with the unique function $g_i = \sum_{k}
\beta_{ik}^0 R(\delta_k, \cdot)$ obtained by solving 
$$\mathbf{g}_i = \mathbb{G} \beta_i^0.$$
This is the minimal norm element whose evaluations coincide with
$\mathbf{g}_i$.  
Since $\mathbb{G}$ is positive definite there are several possible
ways to factorize $\mathbb{G}$ such that $\mathbb{G} = U
U^T$. For the Cholesky factorization $U$ is lower triangular, and for the
spectral decomposition the columns of $U$ are orthogonal. For any such
factorization 
$$\mathbf{g}_i = U \underbrace{U^T \beta_i^0}_{\beta_i} = U \beta_i.$$
Note how the $\beta_i^0$- and thus the $\beta_i$-parameter
representation of the evaluations $\nabla_i \ell^{\text{approx}}(\mathbf{g})(\delta_k)$ can
be read of directly from (\ref{eq:dmllapp1}). We observe that the squared norm of $g_i$ equals 
$$||g_i||^2 = (\beta_i^0)^T \mathbb{G} \beta_i^0 = ||\beta_i||_2^2$$
with $||\cdot||_2$ denoting the ordinary Euclidean norm on
$\mathbb{R}^N$. The parametrization in terms of $\beta_i$ is thus
an isometry from $\mathbb{R}^N$ into $\mathcal{H}$. 
The objective function -- the penalized negative log-likelihood
approximation -- can be computed as 
\begin{equation} \label{eq:ob1}
\ell^{\text{approx}}(U \beta) + \lambda \sum_i ||\beta_i||^2_2
\end{equation}
using  (\ref{eq:mllapp}), and the $\beta_i$-gradient can be computed as
$$U^T \nabla_i^{\beta} \ell^{\text{approx}}(U \beta) + 2\lambda \beta_i,$$
where 
\begin{eqnarray} \nonumber
\nabla_i^{\beta} \ell^{\text{approx}}(\mathbf{g})_k & = & \sum_{l} \phi'(\xi_l) \Delta_l
h_{lik} \\ && \hskip 5mm \label{eq:nablamllapp} - \sum_{l \in I_{\text{jump}}} \frac{\phi'(\xi_l)}{\phi(\xi_l)}
h_{lik}.
\end{eqnarray}
The use of (\ref{eq:mllapp}) and
(\ref{eq:nablamllapp}) -- and (\ref{eq:linpred}) -- requires the
computation of $h_{lik}$. This can either be done on-the-fly (a matrix
free method) or by precomputing the $n \times
  (pr)$-dimensional sparse matrix $\mathbf{H} = (h_{lik})$. In
  practice, an incomplete factorization of
  $\mathbb{G}$ with $U$ an $N \times q$ matrix is used. This reduces the
  number of computations a little, and the transition between the
  $\mathbf{g}_i$ vectors of evaluations and the $\beta_i$-parameters
  becomes numerically more stable. The 
  implementation in \texttt{ppstat} uses the spectral decomposition, and $q$ is
determined by a threshold on the size of the eigenvalues for
$\mathbb{G}$ relative to the largest eigenvalue. The default choice on
the threshold in \texttt{ppstat} is $10^{-8}$.

\subsection{The basis approximation} 

Choose a set of basis functions $B_1, \ldots, B_q$
  such that 
$$\text{span}\{B_1, \ldots, B_q\} \subseteq \text{span}\{R(\delta_0, \cdot),
\ldots, R(\delta_{N-1}, \cdot)\}.$$ 
Precompute the $n \times q$ \emph{model matrices} $\mathbf{Z}^i$ of basis
filters 
$$\mathbf{Z}^i_{lj} = \sum_{k} h_{lik} B_j(\delta_k).$$ 
With $g_i = \sum_{j} \beta_{ij}^0  B_j$, 
the $n$-dimensional linear predictor is given as $\xi = \sum_{i}
\mathbf{Z}^i \beta^0_i$, and $\ell^{\text{approx}}(\beta^0)$ can be computed using
(\ref{eq:mllapp}). The Gram matrix, $\mathbb{G}$, is given by $\mathbb{G}_{kl} = \langle B_k,
B_l \rangle$, and we let $\mathbb{G} = VV^T$. In terms of the
parametrization $\beta_i = V^T \beta_i^0$ we find that 
$$||g_i||^2 = (\beta^0_i)^T \mathbb{G} \beta^0_i = ||\beta_i||^2_2,$$
thus $\beta_i$ provides an isometric parametrization from
$\mathbb{R}^q$ into $\mathcal{H}$. 
The objective function becomes
\begin{equation} \label{eq:ob2}
\ell^{\text{approx}}(V^{-1} \beta) + \lambda \sum_{i} ||\beta_i||_2^2, 
\end{equation}
and the gradient is  
{\small
\begin{eqnarray*} 
&&  \sum_{l} \phi'(\xi_l) \Delta_l (\mathbf{Z}^i_{l} V^{-1})^T  - 
\sum_{l \in I_{\text{jump}}} \frac{\phi'(\xi_l)}{\phi(\xi_l)}
(\mathbf{Z}^i_{l} V^{-1})^T  + 2 \lambda \beta_i \\ 
& = & (V^{-1})^T \left(\sum_{l} \phi'(\xi_l) \Delta_l (\mathbf{Z}^i_{l})^T  - 
\sum_{l \in I_{\text{jump}}} \frac{\phi'(\xi_l)}{\phi(\xi_l)}
(\mathbf{Z}^i_{l})^T\right)  + 2\lambda \beta_i \\
& = & (V^{-1})^T \nabla_{i}^0 l_t^{\text{approx}}(V^{-1} \beta) +  2\lambda \beta_i,
\end{eqnarray*} 
}
where 
{\small 
\begin{eqnarray*}
\nabla_{i}^0 l_t^{\text{approx}}(\beta^0) & = &  \sum_{l} \phi'(\xi_l)
\Delta_l (\mathbf{Z}^i_{l})^T -
\sum_{l \in I_{\text{jump}}} \frac{\phi'(\xi_l)}{\phi(\xi_l)}
(\mathbf{Z}^i_{l})^T
\end{eqnarray*}
}
is the gradient in the $\beta^0_i$ parametrization.

\begin{figure*}
\begin{center}
\begin{tabular}{ccc}
  \includegraphics[width=0.3\textwidth]{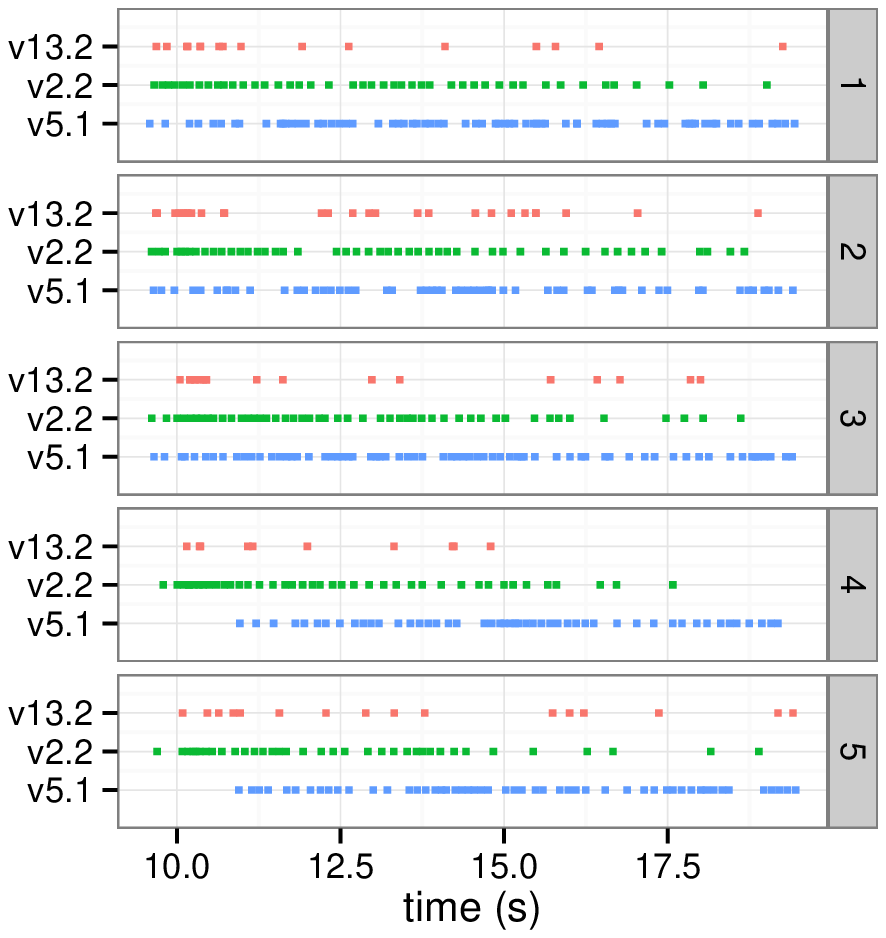} &
  \includegraphics[width=0.3\textwidth]{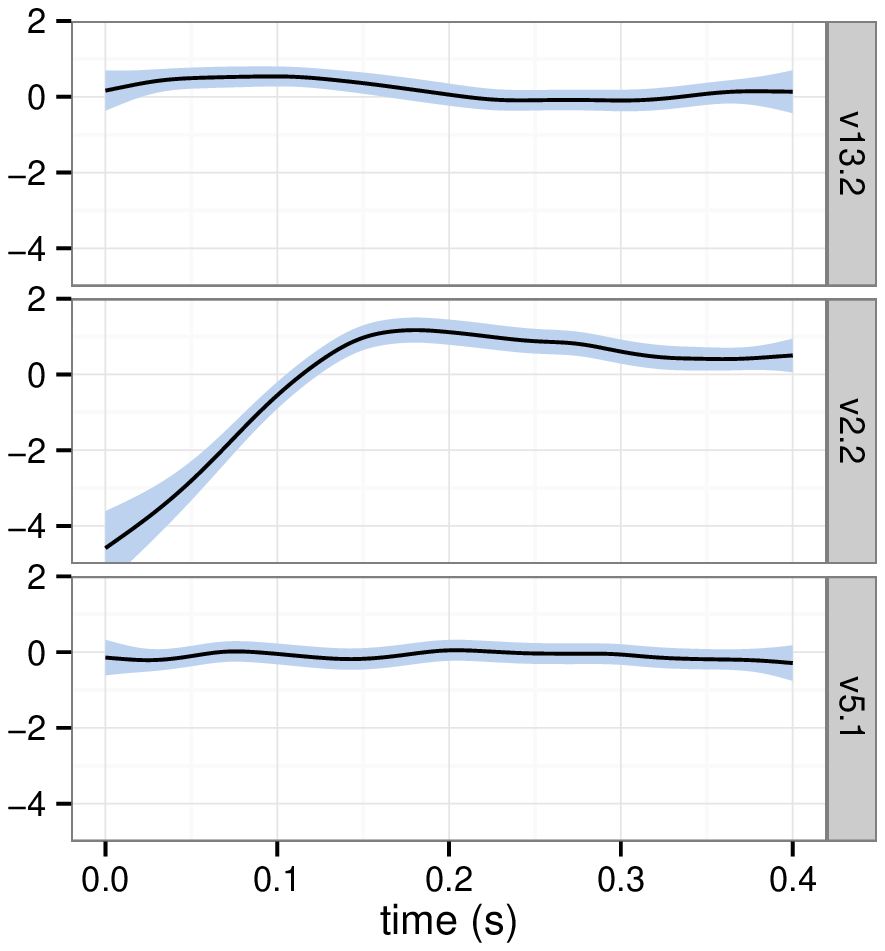} &
  \includegraphics[width=0.3\textwidth]{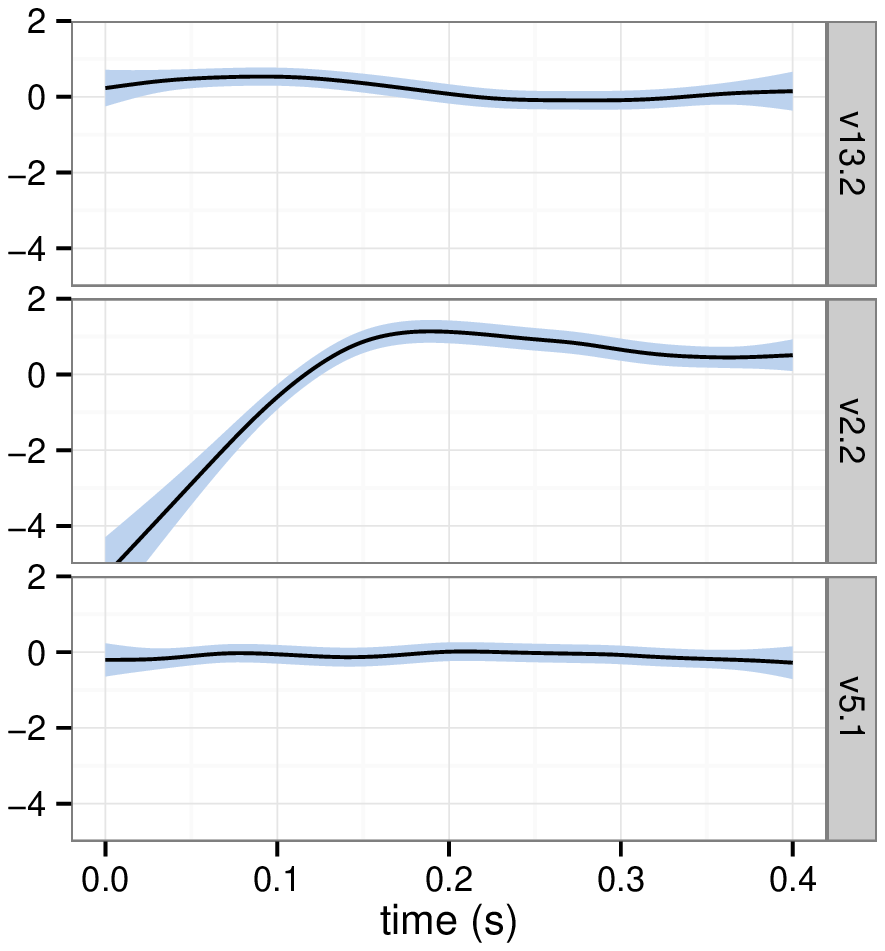} 
\end{tabular}
\end{center}
\caption{Left: Data example consisting of three spike tracks from five independent
  multichannel measurements of turtle spinal neurons during a stimulation
period. Middle: Estimates of the $h_i$'s in the model of \texttt{v2.2} using
the direct approximation to minimize (\ref{eq:ob1}) with $n = 20,609$, $N = 200$ and $q
= 33$. The value of $\lambda = 0.125$ was chosen by minimizing TIC. The
baseline estimate is $\hat{\beta}_0 = 1.04$. 
Right: Similar estimates of the $h_i$'s using a
B-spline basis expansion to minimize (\ref{eq:ob2}) with $q = 33$, and
with $\lambda = 2$ chosen by minimizing TIC. The baseline estimate
is $\hat{\beta}_0 = 1.05$. The point-wise approximate 95\%
confidence intervals were obtained using a sandwich estimator of the
asymptotic variance.}
\label{fig:datafit}       
\end{figure*}

\section{Statistical theory}

In the preceding sections we reduced the infinite dimensional
optimization problem to an approximate finite dimensional one, and we
introduced isometric para-metrizations in terms of a finite
dimensional parameter $\beta$. In this section we discuss the
statistical theory for the case where $X_s(g)
= X_s^T \beta$ for a vector process $X_s$, and the estimator of $\beta$ is obtained by penalized
maximum-likelihood estimation using the penalty $\lambda
||\beta||_2^2$.  If $\phi(X_s^T \beta)$ is the intensity of the
counting process, the process $M_s = N_s - \int_0^s \phi(X_u^T \beta)
\, \mathrm{d} u$ is a local martingale. The derivative of the
negative log-likelihood can be expressed as a stochastic integral
w.r.t. $M$, 
\begin{eqnarray}
D_{\beta} \ell(\beta) & = &  \int_0^t \phi'(X_s^T \beta)  X_s^T \,
  \mathrm{d} s - \int_0^t  \frac{\phi'(X_s^T \beta)}{\phi(X_s^T
  \beta)}  X_s^T \, \mathrm{d}N_s \nonumber \\
& = & - \int_0^t \frac{\phi'(X_s^T \beta)}{\phi(X_s^T \beta)}  X_s^T \,
\mathrm{d} M_s. \label{eq:dell}
\end{eqnarray}
Since the angle bracket of the local martingale $M$ is $\langle M
\rangle_t = \int_0^t \phi(X_s^T \beta) \, \mathrm{d} s$, cf. (2.4.2)
in \cite{AndersenBorganGillKeiding:1993}, it follows from (2.3.7) in
\cite{AndersenBorganGillKeiding:1993} that the angle bracket of the vector process
(\ref{eq:dell}) is 
$$ \int_0^t  X_s X_s^T \frac{\phi'(X_s^T \beta)^2}{\phi(X_s^T \beta)^2} \,
\mathrm{d} \langle M_s \rangle =  \int_0^t X_sX_s^T
  \frac{\phi'(X_s^T\beta)^2}{\phi(X_s^T \beta)} \,
  \mathrm{d}s.$$
Under sufficient integrability conditions, see e.g. Proposition 4.6.2
in \cite{Jacobsen:2006}, the mean of the angle bracket equals
the covariance matrix $K$ of (\ref{eq:dell}), which is the 
Fisher information matrix. This suggests the estimator 
$$\hat{K} =  \int_0^t X_sX_s^T
  \frac{\phi'(X_s^T\hat{\beta})^2}{\phi(X_s^T \hat{\beta})} \,
  \mathrm{d}s$$
of the Fisher information, which can be computed using the same time discretization as otherwise
used for likelihood and gradient computations. 

From (\ref{eq:dell}) we find the penalized likelihood estimating equation in the
$\beta$-parametrization
\begin{equation} \label{eq:penlikeeq}
\Psi(\beta) := - \int_0^t \frac{\phi'(X_s^T \beta)}{\phi(X_s^T \beta)}  X_s^T \,
\mathrm{d} M_s + 2 \lambda \beta^T = 0.
\end{equation}
The covariance matrix of $\Psi(\beta)$ defined above coincides with the
Fisher information matrix $K$, and the mean of its derivative is 
$$J := E D_{\beta} \Psi(\beta) = K + 2 \lambda I.$$ 
A corresponding estimator of $J$ is $\hat{J} = \hat{K} + 2 \lambda
I$. If $\hat{\beta}_{\lambda}$ solves (\ref{eq:penlikeeq}), a Taylor
expansion of $\Psi(\hat{\beta}_{\lambda})$ around $\beta$ yields 
$$0 = \Psi(\hat{\beta}_{\lambda}) \simeq \Psi(\beta) + D_{\beta}
\Psi(\beta) (\hat{\beta}_{\lambda} - \beta),$$
which implies the approximation
$$\hat{\beta}_{\lambda} - \beta \simeq - D_{\beta}
\Psi(\beta)^{-1} \Psi(\beta) \simeq - J^{-1} D_{\beta}\ell(\beta) - 2
\lambda J^{-1} \beta.$$
If, moreover, the distributional approximation 
$$D_{\beta}\ell(\beta) \overset{\text{approx}}{\sim} \mathcal{N}(0,
K)$$ can be justified, these heuristic derivations suggest that 
$$
\hat{\beta}_{\lambda} \overset{\text{approx}}{\sim} \mathcal{N}(\beta -
2 \lambda J^{-1}\beta,  J^{-1} K J^{-1}).$$
We will not attempt to establish sufficient technical conditions in an asymptotic
framework to rigorously justify this approximation of the distribution
of $\hat{\beta}_{\lambda}$, but see \cite{vanderVaart:1998} for a
treatment of standard asymptotic theory, and Chapter VI in
\cite{AndersenBorganGillKeiding:1993} for a treatment in a
counting process framework. We use the distributional approximation to compute pointwise confidence
intervals based on the sandwich estimator 
$$\hat{J}^{-1} \hat{K} \hat{J}^{-1}$$
of the approximate covariance matrix of $\hat{\beta}_{\lambda}$. 

With $\ell(\hat{\beta}_{\lambda})$ the negative log-likelihood in the
penalized estimator we also introduce Takeuchi's information criterion
$$\textrm{TIC} = \ell(\hat{\beta}_{\lambda}) +
\mathrm{tr}(\hat{J}^{-1} \hat{K}),$$
see Chapter 2 in \cite{Claeskens:2008}. The penalization
parameter $\lambda$ can be chosen by minimizing $\textrm{TIC}$.

\section{Results}

We investigated the use of both the direct approximation and the basis
expansion using cubic B-spline basis functions on a test data
set of neuron spike times. The data set consisted of multichannel
measurements of spinal neurons from a turtle. The measurements were
replicated 5 times and each time the spike activity was recorded 
over a period of 40 seconds. A 10 seconds stimulation was given within
the observation window. We used the spike times for 3 neurons labeled
\texttt{v2.2}, \texttt{v13.2} and \texttt{v5.1}  
during the stimulation period, see Figure \ref{fig:datafit}

The likelihood and gradient algorithms are
implemented in the R package \texttt{ppstat}, which supports optimization of
the objective function via the R function \texttt{optim} using the
BFGS-algorithm. The \texttt{ppstat} package offers a formula based model
specification with an interface familiar from \texttt{glm}. The direct
approximation is
implemented via the \texttt{ppKernel} function, and the basis expansion is
implemented via the \texttt{ppSmooth} function. A typical call has
the form

{\small
\begin{verbatim}
ppKernel(v2.2 ~ k(v13.2) + k(v2.2) + k(v5.1),
                     data = spikeData,
                     family = Hawkes("log"),
                     support = 0.4,
                     lambda = 0.125
)
\end{verbatim}
}

\noindent which will include a baseline parameter in addition to the
three nonparametric filter functions. The data set contained in the
object \texttt{spikeData} must be of class \texttt{MarkedPointProcess}
from the supporting R package \texttt{processdata}. The grid of $n$
time points is determined when the \texttt{MarkedPointProcess}
object is constructed. The choice of $\phi$ is specified as an ``inverse link
function'' -- being \texttt{"log"} in the call above. That is, in the
call above, $\phi(x) = e^x$. 

Figure \ref{fig:datafit} shows the estimated $h_i$'s obtained with
$\phi(x) = e^x$, $A = 0.4$, $n = 20,609$, $N = 200$ and $q=33$ and
using either the direct approximation with the Sobolev kernel or the basis
expansion with a B-spline basis.  The estimates were computed by
minimizing (\ref{eq:ob1}) and (\ref{eq:ob2}), respectively. The
Sobolev kernel is the reproducing kernel for the Sobolev Hilbert
space consisting of twice weakly differentiable functions with the
second derivative being square integrable. Its precise form depends on
which inner product is chosen, but for common choices $R(\delta_k,
\cdot)$ is a cubic spline. The resulting model shows that a \texttt{v2.2} spike results in a 
depression of the \texttt{v2.2}-intensity in the first 0.1 seconds after the spike followed by 
an elevation of the \texttt{v2.2}-intensity. A \texttt{v13.2} spike appears to result in a 
small but significant elevation of the \texttt{v2.2}-intensity, whereas a \texttt{v5.1}
spike appears to have no significant effect on the
\texttt{v2.2}-intensity.

\begin{figure*}
\begin{tabular}{c}
  \includegraphics[width=\textwidth]{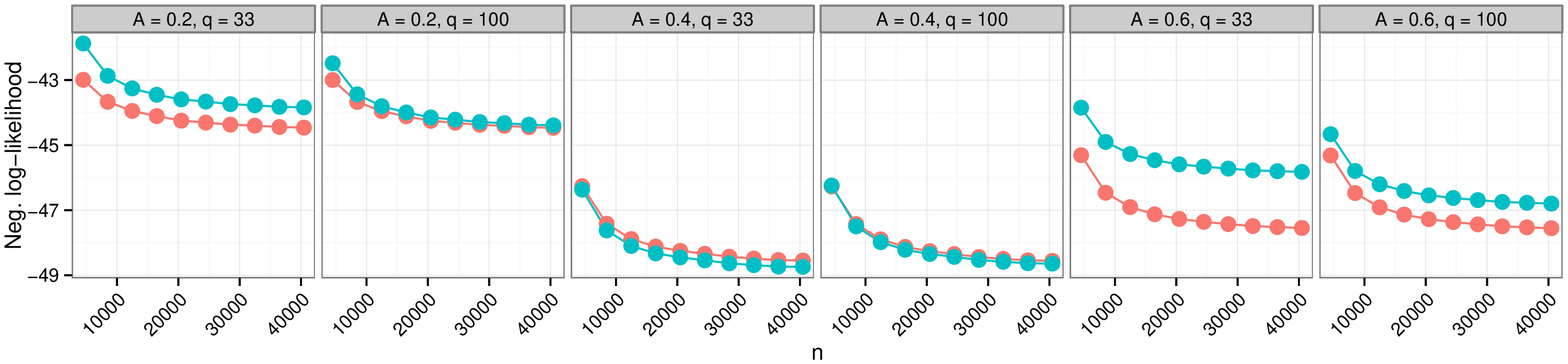} \\
  \includegraphics[width=\textwidth]{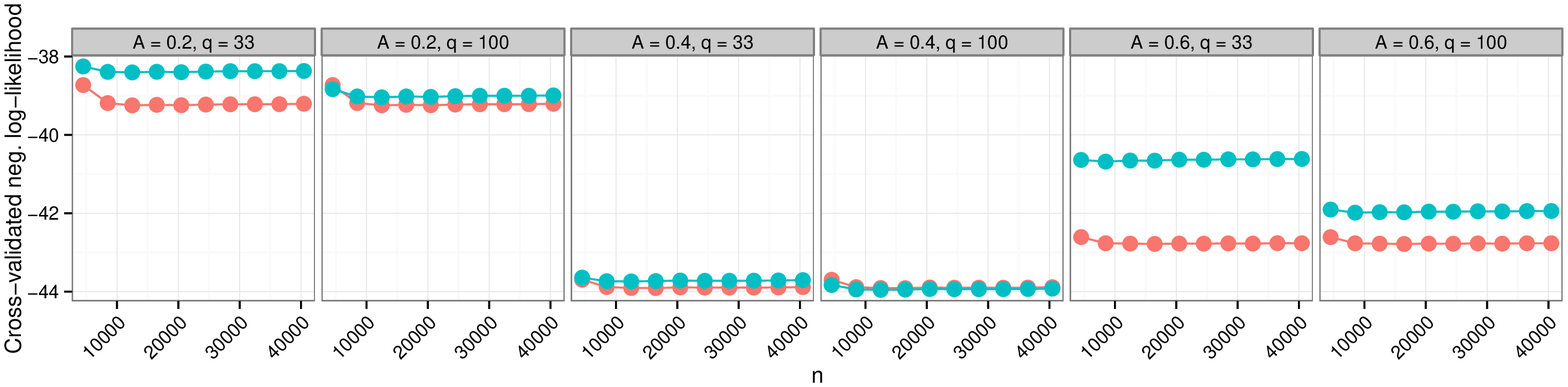} \\
\end{tabular}
\caption{Top: Average negative log-likelihood on the training data for
  the direct approximation (\textcolor{A}{$\CIRCLE$}) and for the
  basis expansion (\textcolor{B}{$\CIRCLE$}) for $N = 400$, $A = 0.2$,
  $0.4$, $0.6$ and $q = 33$, $100$. Bottom: Cross-validated negative
  log-likelihood.  }
\label{fig:accuracy}       
\end{figure*}

The \texttt{ppstat} package supports a
number of different choices of $\phi$. In addition to some familiar
link functions it supports two parametrized classes of
functions. The \texttt{root} class is given as 
$$\phi(x) = \left\{\begin{array}{ll} 
x^{c + 1} & \quad \text{for } x \leq 0 \\
0 & \quad \text{for } x > 0,
\end{array}\right.$$
which for $c = 0$ (the default value) amounts to $\phi(x) = \max\{x,
0\}$. The \texttt{logaffine} class is given as
$$\phi(x) = \left\{\begin{array}{ll} 
e^x & \quad \text{for } x \leq c \\
e^c(x  - c + 1) & \quad \text{for } x > c.
\end{array}\right.$$
They all map $\mathbb{R}$ continuously into $[0, \infty)$. Moreover,
for the \texttt{logaffine} class the $\phi$ function is
continuously differentiable, whereas for the \texttt{root} class this
is only true for $c > 1$. 

The appropriate choice of $\phi$ is not straight forward, and there
are several considerations that need to be taken into account. One
possibility, that we have used, is to optimize the model fit to data. 
However, one must pay attention to the fact that not all combinations of
$\phi$ and linear filters will result in non-exploding point
processes, see \cite{Gjessing:2010}. This will be particularly
problematic if we were to simulate data from the model. It is
difficult to give theoretical results on the non-explosion or 
stability of a point process if $\phi$ grows super linearly, 
see e.g. \cite{Bremaud:1996} where $\phi$ is assumed to be Lipschitz
to establish results on stability of point processes.

We chose to consider the \texttt{logaffine} class and
to optimize over $c$ to achieve the best model fit. Since the optimal
choice of the penalization parameter $\lambda$ may be affected by the
choice of $c$, we minimized TIC over a grid of $c$ and $\lambda$
values. When the memory bound on the linear filters was chosen as $A = 0.4$, the optimal
choice of $c$ was effectively $+ \infty$, meaning that $\phi(x) = e^x$
was optimal.

\begin{figure*}
\begin{tabular}{c}
  \includegraphics[width=\textwidth]{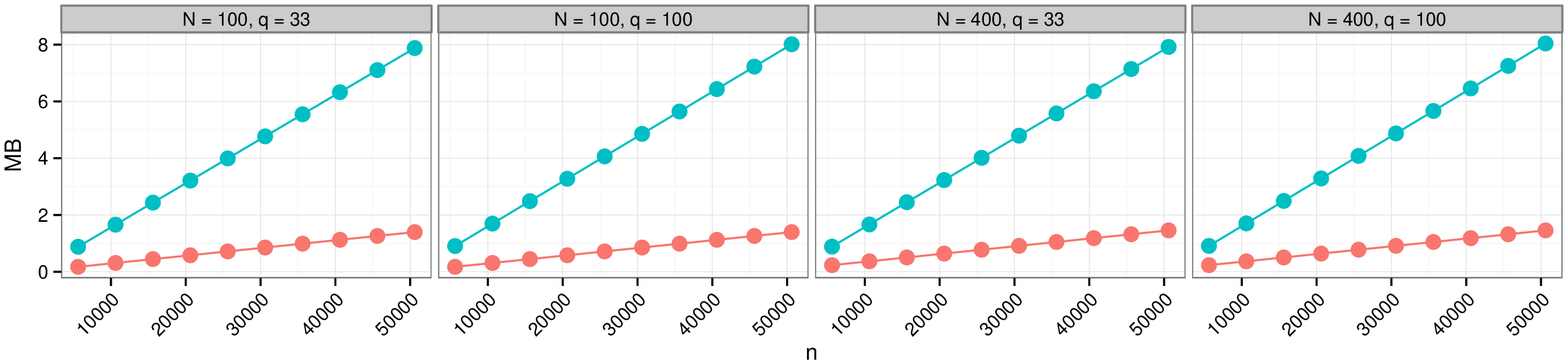} \\
  \includegraphics[width=\textwidth]{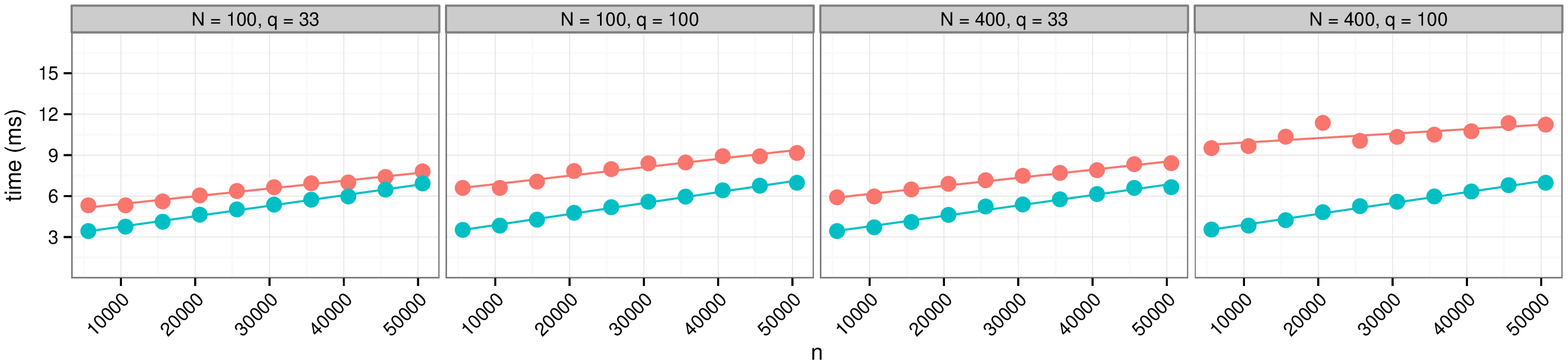} \\
  \includegraphics[width=\textwidth]{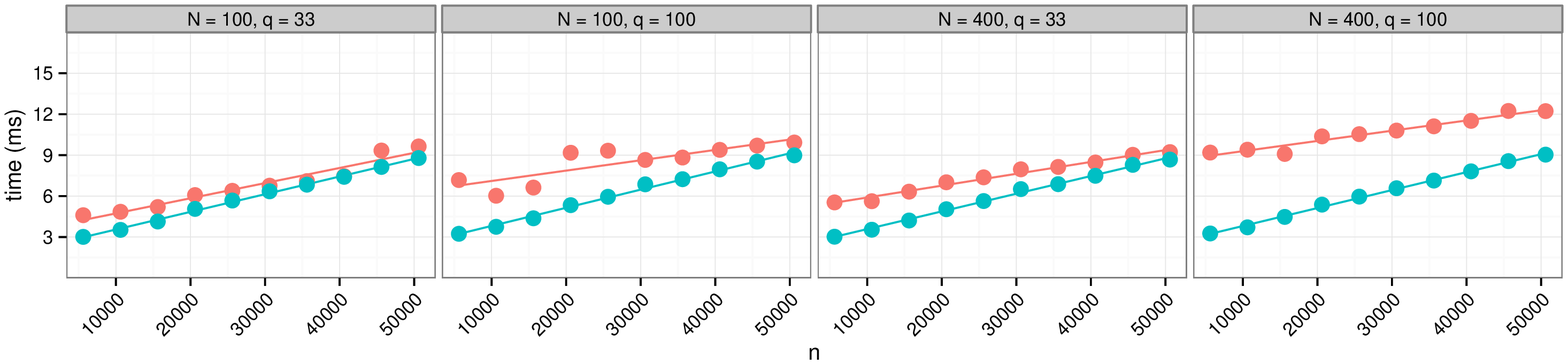} \\
\end{tabular}
\caption{Top: Memory usage for storing the $\mathbf{H}$-matrix for
  the direct approximation (\textcolor{A}{$\CIRCLE$}) and the  $\mathbf{Z}$-matrix
 for basis expansion (\textcolor{B}{$\CIRCLE$}) for $N = 100$, $400$ and $q
 = 33$, $100$. Middle: Log-likelihood computation time. Bottom: Gradient
 computation time.}
\label{fig:memtime}       
\end{figure*}

We then investigated how several of the other model and
approximation choices affected the model fit. The data set consists of 
five replications, and we used cross-validation to iteratively fit the model on
four of the five replications (the training data) and validate it on the last. 
We did this for different combinations of $A$, $n$ and $q$ that
determine the memory bound, the resolution of the time discretization and the dimension
of the actual parameter space. We fixed $N = 400$ for these computations. For the basis expansion the number of B-spline
basis functions was chosen explicitly to be either $q = 33$ or $q =
100$. For the direct approximation the choice of threshold on the spectrum was 
tuned to result in $q = 33$ or $q = 100$. The negative log-likelihood per replication was
used as a measure of model fit, and it was in all cases computed using
the finest time discretization (largest $n$). Figure
\ref{fig:accuracy} shows that the fit generally improved on the
training data as $n$ increased, but above $n = 20,000$ the improvement
was small. The cross-validated negative log-likelihood was, however,
almost unaffected by the choice of $n$ used for fitting the
model. Only the coarsest discretization resulted in a slightly worse 
fit.  Figure \ref{fig:accuracy} also shows that among the three
choices of $A$, $A = 0.4$ was best both in terms of model fit on the
training data and in the cross-validation. Finally, the choice of $q$
did not play a role for the direct approximation in terms of model
fit, whereas for the basis expansion the use of more basis functions
resulted in a slightly better model fit, in particular for $A = 0.6$.

We then investigated the memory usage and the computation times of 
the direct approximation and the basis expansion. The memory usage was obtained using the R function
\texttt{object.size} and the computation times were computed as the
average of 80 replicated likelihood or gradient evaluations. The
interest was on how they scale with the numbers $n$, $N$ and $q$. The
implementation relies on precomputation of the $\mathbf{H}$ or
$\mathbf{Z}$ matrices, which are stored as sparse matrices as
implemented in the R package \texttt{Matrix}.  Note that for the 
basis approximation the choice of $N$ only affects the precomputation of the model
matrix and not the likelihood and gradient computations. 

Figure \ref{fig:memtime} shows that the basis expansion used
more memory for storing $\mathbf{Z}$, and that the memory usage as a
function of $n$ had a somewhat larger slope than for the direct
approximation. We should note that the memory usage for neither of the methods 
showed a noticeable dependence upon $N$ or $q$. Storing the matrices
as non-sparse matrices the $\mathbf{Z}$-matrix required
119 MB and the $\mathbf{H}$-matrix required 465 MB for $n = 50,000$, $N
= 400$ and $q = 100$. In comparison, the sparse versions required 8 MB and 1.5
MB, respectively. 

Figure \ref{fig:memtime} shows, furthermore, that likelihood
and gradient computations were generally faster when the basis
expansion was used. More
importantly, Figure \ref{fig:memtime} shows that computation time 
for the direct approximation depended upon $N$ as well as $q$, and that the
computation times for the basis expansion, using the B-spline basis, 
were remarkably independent of $q$. Note also that the slope on the
computation times, as a function of $n$, is slightly smaller for the direct approximation
than for the basis expansion. The dominating contribution to the
computations are the matrix-vector products $\mathbf{H} (U \beta)$ and 
 $\mathbf{Z} (V^{-1} \beta)$. The former scales approximately like $a
 N p n  + p^2 N q$ and the latter like $b q p n + p^2 q^2$, where $a$ and $b$ are
 the fraction of non-zero entries in the matrices. Our results reflect 
 that $aN < bq$.

\section{Discussion}

The two approximations considered in this paper
differ in terms of what is precomputed. Computing the matrix
$\mathbf{H}$ upfront as in the direct approximation
should require only a fraction of the memory required for storing the
$\mathbf{Z}$-matrices. This was confirmed by our
implementation. We also showed that the storage
requirements for the direct approximation did not depend
noticeably on the number $N$ of $\delta$-grid points when $\mathbf{H}$ is stored as
a sparse matrix. The tradeoff is an increased computation
time, which depends on the resolution determined by $N$ and
$q$. 

The storage requirements for $\mathbf{Z}$ can easily become
prohibitively large. A choice of basis functions with local
support, such as B-splines used here, can compensate
partly for this. It is unlikely that it is useful to
precompute $\mathbf{Z}^i V^{-1}$, as this will destroy the
computational benefits of the basis with local support.

For the basis expansion it is possible to precompute the model matrix in a slightly
different and more direct way. Instead of precomputing the $q \times N$ basis function
evaluations $B_j(\delta_k)$ we can compute $\mathbf{Z}_{lj}^i$
directly as 
$$\mathbf{Z}_{lj}^i = \sum_{k: t_l - A < \sigma_k^i < t_i}  B_{j}(t_l -
\sigma_k^i).$$
This may be more accurate but since $n \gg N$ in typical applications
this comes at the cost of many more basis function
evaluations. Whether this is critical in terms of the time to compute
$\mathbf{Z}^i$ depends upon how costly a single basis function
evaluation is relative to the computation of the $h_{lik}$'s. We have
not presented data on the computational costs of the precomputations,
but they were observed to be small compared to the costs of the actual
optimization. 

We observed that the fitted models obtained by either the direct
approximation using the Sobolev kernel or the B-spline basis expansion
were almost identical. This is not surprising given the fact that 
$R(\delta_k, \cdot)$ is a cubic spline. In the actual implementation 
there are minor differences -- for the B-spline expansion the 
linear part is, for instance, not penalized whereas all parts of the kernel fit is 
penalized. In conclusion, the B-spline basis expansion is
currently to be preferred if the storage requirements can be met. 
The implementation of the direct approximation
does, however, offer an easy way to use alternative kernels and thus
alternative reproducing kernel Hilbert spaces.

We illustrated the general methods and the implementation using neuron network
data. In practice, there are many model choices to be made besides the
choice of appropriate discretizations. We have shown how some of these
choices, e.g. the choice of $\phi$ and the choice of $\lambda$,  
can be made by minimizing TIC. The choice of $A$ can be made
on a data driven basis in a similar way. Neuron network activity is just one example of a multivariate
interacting dynamical system that is driven by discrete events. Other 
examples include high-frequency trading of multiple financial assets, see
\cite{Hautsch:2004}, and chemical reaction networks as discussed in
\cite{Anderson:2011} and \cite{Bowsher:2010}. The Markovian linear Hawkes model
(\ref{eq:expfilt}) was also considered in Chapter 7 in
\cite{Hautsch:2004}, and the typical models of chemical reactions are
Markovian multitype birth-death processes. Markovian models 
are often computationally advantageous, as they offer more 
efficient intensity and thus likelihood computations. With the
implementation in the R package \texttt{ppstat} we have made more
flexible yet computationally tractable 
nonparametric and non-Markovian models available.

\section{Proof of Proposition \ref{prop:grad}} \label{sec:proof}

Function evaluations are represented in terms of
the kernel by inner products as given by (\ref{eq:rep1}). This gives that  
\begin{equation} 
X_s(g) = \sum_{i=1}^p \left\langle  \sum_{j: \sigma_j^i < s} R(s -
\sigma_j^i, \cdot), \ g_i \right\rangle. \label{eq:rep}
\end{equation}

If $\psi$ is a continuously differentiable function we find that 
\begin{eqnarray*}
&& \frac{\psi(X_s(g + \varepsilon h)) - \psi(X_s(g))}{\varepsilon} \\
&& \hskip 1mm =
\frac{\psi(X_s(g) + \varepsilon X_s(h)) - \psi(X_s(g))}{\varepsilon}
\longrightarrow \psi'(X_s(g)) X_s(h)
\end{eqnarray*}
for $\varepsilon \rightarrow 0$. This is clearly a continuous linear
functional.  Using (\ref{eq:rep}) and differentiating only w.r.t. the $i$'th
coordinate of $g$ we find that the corresponding gradient in
$\mathcal{H}$ is 
$$\nabla_i \psi(X_s(g)) =  \psi'(X_s(g)) \sum_{j: \sigma_j^i < s} R(s -
\sigma_j^i, \cdot).$$
Taking $\psi = \log \phi$ this yields the gradient of the second term
in the negative log-likelihood, $\sum_{k=1}^{N_t}
\log \phi(X_{\tau_k}(g))$, directly. For the first term we take $\psi
= \phi$, but we need to
ensure that we can interchange the order of 
integration and differentiation. To this end the following norm bound on
$\nabla_i \phi(X_s(g))$ is useful
\begin{eqnarray*}
||\nabla_i \phi(X_s(g))|| & \leq & |\phi'(X_s(g))|  \sum_{j: \sigma_j^i < s} ||R(s -
\sigma_j^i, \cdot)|| \\
& \leq & C_t N^i_t \sup_{s \in [0,t]} \sqrt{R(s, s)} < \infty. 
\end{eqnarray*}
Here $C_t = \sup_{s \in [0,t]} |\phi'(X_s(g))|$ is finite because
$X_s(g)$ is continuous in $s$ and $\phi'$ is assumed continuous. We
have also used that $||R(s - \sigma_j^i, \cdot)||^2 = R(s -
\sigma_j^i, s - \sigma_j^i)$ and the fact that $R$ is continuous to
conclude that the bound is finite. The bound shows that 
$$\sum_{j} \int_{\sigma_j^i}^t \phi'(X_{s}(g))
R(s - \sigma_j^i, \cdot) \, \mathrm{d} s$$
is an element in $\mathcal{H}$, and the required interchange of integration and
differentiation is justified by the bound. This completes the proof. 
\qed

\begin{acknowledgements}
The neuron spike data were provided by Associate Professor, Rune
W. Berg, Department of Neuroscience and pharmacology, University of
Copenhagen.  
\end{acknowledgements}


\end{document}